# Influence of Indium-Percentage Variation on Dynamical Characteristics of In$_x$Ga$_{1-x}$As/GaAs(001) Quantum Dot Lasers

*Mahdi Ahmadi Borji[*] and Esfandiar Rajaei[**]*

***Department of Physics, the University of Guilan, Namjoo Street, Rasht, Iran***

*[*] Email: Mehdi.p83@gmail.com, [**]Corresponding author, E-mail: Raf404@guilan.ac.ir*

*Abstract:*

*The influence of indium percentage on dynamical characteristics of In$_x$Ga$_{1-x}$As/GaAs(001) quantum dot lasers (QDLs) is investigated. Energy levels of self-organized truncated-cone-shape QDs are calculated by means of the eight-band k.p model, and their dependence to indium percentage is surveyed. Then, by presenting a three-level model and numerical solution of the resulting rate equations, laser properties are determined. Our results show that inclusion of more indium gives rise in the reduced energy gap and electron-hole recombination energy. Moreover, lasing process for both Ground State (GS) and Excited States (ES) sound to be sensitive to indium percentage. It is shown that rise of indium percentage at fixed injected current results in the increased ES turn-on delay and GS photon number and 3dB modulation bandwidth, and decreased ES photon number, GS turn on delay, amplitude of relaxation oscillations, output power, and ES 3dB modulation bandwidth; but has no effect on threshold current and laser gain. At last, we find an optimized cavity length which was likely to be independent of indium percentage.*

## I. INTRODUCTION

Study of semiconductor structures enables to control their parameters such as energy levels, energy gap, and the band structures. Engineering these factors leads to novel high-performance devices such as many types of lasers, each of which being suitable for a special application. Semiconductor nanostructures include quantum wells (QWs), quantum wires (QWRs), and quantum dots (QDs) in which carriers are restricted to one, two, or three dimensions respectively. Quantum confined semiconductor nanostructures have been the focus of many researches due to their optical and electronic properties arising from quantum confinement of electrons and holes [1-4]. In semiconductor quantum dots, carriers are restricted to a three-dimensional space smaller than de Broglie wavelength. This confinement arises if band gap of the surrounded nanomaterial is less than that of the surrounding semiconductor [5]. Progress in the fabrication of QD-lasers has recently attracted a huge attention to the application of quantum systems in optoelectronics [6-9].

InAs based QDLs are the most peculiar cases in the local area networks, because of the inherent low threshold current, enhanced modulation speed, and temperature stability [10, 11]. Many research groups attempt to optimize QDLs [6, 12-16]. By changing the compositions, QDs can be engineered, and finding a way to enhance the efficiency of QDLs can be helpful. Many scientists are interested in $In_xGa_{1-x}As$/GaAs QD laser devices due to their interesting and applicable features [4, 12, 13, 17, 18].

It is shown that variation of materials percentages in epitaxial growth of QDs affects on both QD size and density, which directly changes the e-h recombination energies [19-22]. Some theoretical and experimental works elaborated to find the relationship between QD size and mismatch of the QD with the substrate [23, 24]. Despite the relevance of quantum dot features and the result in lasing, there is in the literature a lack of simulations elaborating to find a way to connect the parameters of the structure of QD with the final laser properties. Most of the investigations conducted in this field have focused on only one of energy level change by indium content or lasing process, and they have not investigated the direct effect of stoichiometric percentage on lasing, and hardly a paper can be seen studying the stoichiometry effect in laser applications. Indium percentages of 0.4 and 0.5 have been used in laser devices [25-27]. Strain leads to formation of different size of QDs. However, while changing the indium percentage, we assume the QD size to remain fixed, since the change of indium percentage is restricted to a narrow interval, i.e., 40% to 60% which does not make a remarkable change on QD size.

To date, many quantum solutions of Schrödinger equation are introduced, among which k.p approach with $8 \times 8$ matrix is a good approximation [28, 29]. Mismatch between the lattice constants of substrate and formed QD is dependent to substrate index [30]. We supposed the QDs to grow on (001) index of substrate.

The rest of this article is organized as follows; section II is devoted to numerical calculation of QD energy levels and their behavior in different indium contents by k.p model; in section III-A our model for laser dynamics is introduced, and the results are discussed in III-B; eventually, we make a conclusion in section IV.

**II. ENERGY LEVEL ENGINEERING**

In this research a truncated cone-shape QD is studied. The eight-band k.p approach was employed to solve the Schrödinger equation. Minimized strain is taken into account and the Dirichlet boundary condition for the electrostatic potential is considered. For a better performance of the simulation, a wide region is solved semi-classical but inside the dot the solution is performed with quantum mechanical approach [31].

Fig. 1 illustrates the profile of a truncated-conic InGaAs QD with bottom and top radiuses of 14nm and 8.4nm respectively, height of 10.5nm (i.e., $H = \frac{3D}{8}$ near the value ratio taken in the experimental work in [25]), and a wetting layer of thickness 1.2nm grown on the (001) surface of GaAs. Temperature is assumed to be $T = 300\ K$.

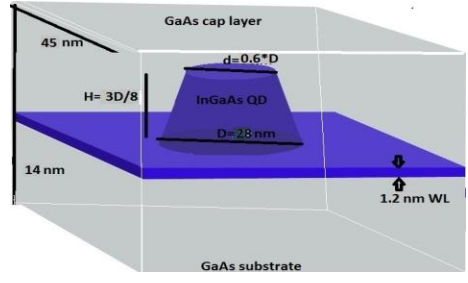

*Fig. 1: Schematics of a truncated cone-shape InGaAs QD with bottom and top diameters of 28nm and 8.4nm respectively, height of 10.5 nm, and a wetting layer of height 1.2nm.*

The parameters related to the bulk materials applied in this paper are presented in Table 1 [32-34].

| Parameters used | GaAs | InAs |
|---|---|---|
| Band gap (0K) | 1.424 eV | 0.417 eV |
| lattice constant | 5.65325 Å | 6.0583 Å |
| Effective electron mass (Γ) | $0.067m_o$ | $0.026m_o$ |
| Effective heavy hole mass | $0.5m_o$ | $0.41m_o$ |
| Nearest neighbor distance (300K) | 0.2448 nm | 0.262 nm |

*Table 1: Parameters included in the model*

At 300K the parameters are computed for $In_xGa_{1-x}As$ as follows:

Lattice constant [35]:

$$a = (6.0583 - 0.405(1 - x)) \, Å \qquad (1)$$

Effective electron mass [36]:

$$m_e = (0.023 + 0.037(1 - x) + 0.003(1 - x)^2)m_o \qquad (2)$$

Effective heavy hole mass [37]:

$$m_h = (0.41 + 0.1(1 - x))m_o \qquad (3)$$

In Figs. 2(a)-2(c) snapshots of conduction (Γ) and valence (Heavy-Hole (HH)) band-edges are depicted as a function of distance in z-direction for different ratios of indium content; also, three first energy states for electrons and holes are inserted too. Dimensional confinement, as it is seen, has separated the energy bands to atomic-like levels. Inclusion of more amount of indium, as it is observed by comparison of 2(a-c), has lead to closer electron-hole levels and smaller energy gap. Ground State (GS), and the Excited States (i.e., $ES_1$ and $ES_2$) as well as the relating values of stimulated recombination energies are schematically included to the figures which are changed by indium percentage.

Energy gap in T=300K for bulk InAs and GaAs are respectively 0.36eV and 1.43eV [38]. It has been shown that energy gap of $In_xGa_{1-x}As$ is sensitive to the indium concentration too. The band gap here gets narrower by increase of indium content. For bulk materials, it is reported in reference [39] that variation of indium percentage in $Ga_xIn_{1-x}As$ affects the band gap with the following functionality:

$$E_g(x)|_{T=300K} = (0.36 + 0.63x + 0.43x^2) \quad (4)$$

which shows a reduction in the band-gap by increase of indium content; Similarly, our figure show that band-gap, which is the difference of conduction band minimum and valence band maximum is reduced while indium percentage is enhanced.

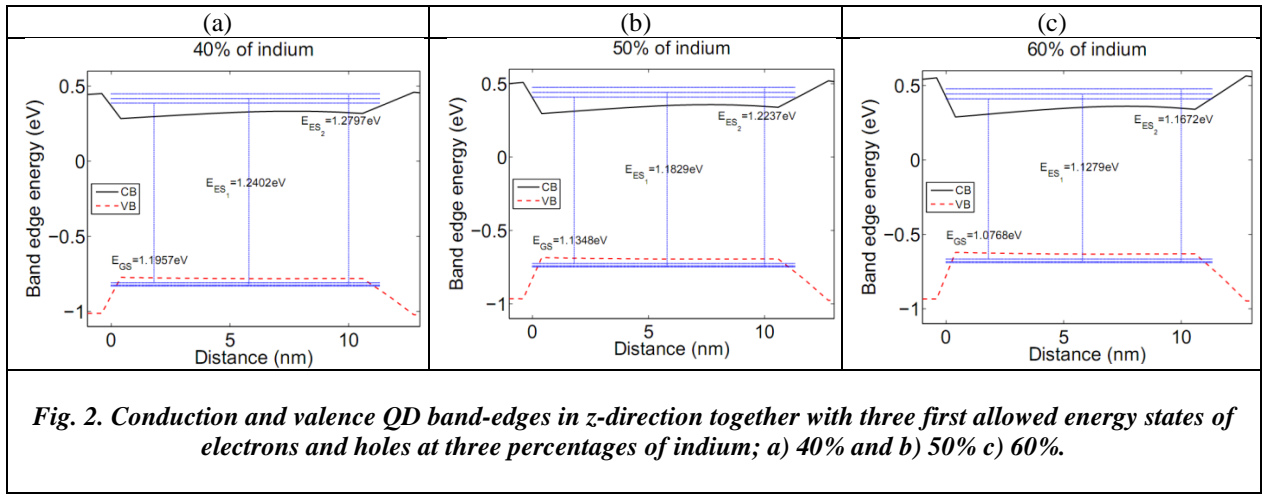

Fig. 2. Conduction and valence QD band-edges in z-direction together with three first allowed energy states of electrons and holes at three percentages of indium; a) 40% and b) 50% c) 60%.

### III. LASING PROCESS

**III-A. Theoretical model**

Schematic energy diagram of a three-level QD laser with GS, $ES_1, ES_2$, and WL levels is illustrated in Fig. 3. Current pumped into the device is captured from WL into the QD's higher level $ES_2$ and can also relax into lower levels or directly recombine with holes and emit photons via stimulated emission. Recombination can happen thus from all three discrete levels. Each of capture, escape, relaxation or recombination with holes takes a while. The time is dependent on the probability of occupation of origin and destination levels, requirements of Pauli Exclusion Principle, phonon bottleneck effect, etc; all the times are very short (~ps-ns) [40]. By taking into account all the transitions shown in the figure, and considering the homogeneous and inhomogeneous broadening, and nonlinear gain, the rate equations can be written as follow:

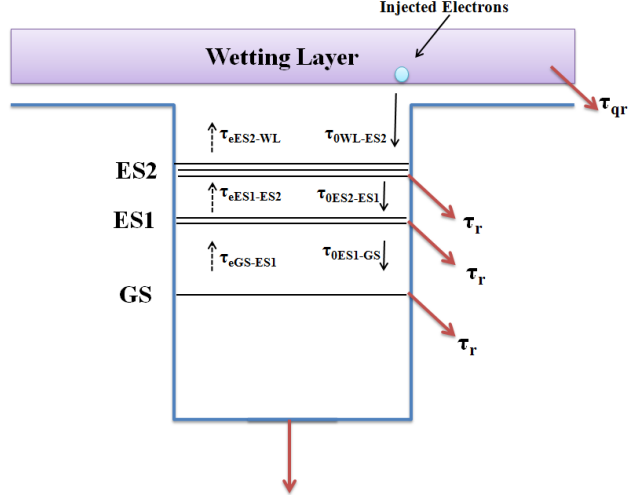

*Fig. 3: Schematics of carrier transitions in three-level model of a QD-laser.*

$$\dot{N}_{WL} = \eta_i \frac{I}{e} - \frac{N_{WL}}{\tau_{qr}} + \frac{N_{ES2}}{\tau_{e_{ES2}}} - \frac{N_{WL}}{\tau_c}(1 - f_{ES2}) \tag{5}$$

$$\dot{N}_{ES2} = -\frac{N_{ES2}}{\tau_r} - \frac{N_{ES2}}{\tau_{e_{ES2}}} - \Gamma v_g K_{ES2} \frac{S_{ES2}(2f_{ES2}-1)}{1+\epsilon_{ES2}S_{ES2}} + \frac{N_{WL}}{\tau_c}(1 - f_{ES2}) - \frac{N_{ES2}(1-f_{ES1})}{\tau_{0ES2ES1}} + \frac{N_{ES1}}{\tau_{e_{ES1}}}(1 - f_{ES2}) \tag{6}$$

$$\dot{N}_{ES1} = -\frac{N_{ES1}}{\tau_r} - \Gamma v_g K_{ES1} \frac{S_{ES1}(2f_{ES1}-1)}{1+\epsilon_{ES1}S_{ES1}} + \frac{N_{GS}}{\tau_{e_{GS}}}(1 - f_{ES1}) - \frac{N_{ES1}(1-f_{GS})}{\tau_{0ES1GS}} + \frac{N_{ES2}(1-f_{ES1})}{\tau_{0ES2ES1}} - \frac{N_{ES1}(1-f_{ES2})}{\tau_{e_{ES1}}} \tag{7}$$

$$\dot{N}_{GS} = -\frac{N_{GS}}{\tau_r} - \Gamma v_g K_{GS} \frac{S_{GS}(2f_{GS}-1)}{1+\epsilon_{GS}S_{GS}} - \frac{N_{GS}}{\tau_{e_{GS}}}(1 - f_{ES1}) + \frac{N_{ES1}}{\tau_{0ES1GS}}(1 - f_{GS}) \tag{8}$$

$$\dot{S}_{ES2} = -\frac{S_{ES2}}{\tau_s} + \Gamma v_g K_{ES2} \frac{S_{ES2}(2f_{ES2}-1)}{1+\epsilon_{ES2}S_{ES2}} + \beta_{sp} \frac{N_{ES2}}{\tau_{sp}} \tag{9}$$

$$\dot{S}_{ES1} = -\frac{S_{ES1}}{\tau_s} + \Gamma v_g K_{ES1} \frac{S_{ES1}(2f_{ES1}-1)}{1+\epsilon_{ES1}S_{ES1}} + \beta_{sp} \frac{N_{ES1}}{\tau_{sp}} \tag{10}$$

$$\dot{S}_{GS} = -\frac{S_{GS}}{\tau_s} + \Gamma v_g K_{GS} \frac{S_{GS}(2f_{GS}-1)}{1+\epsilon_{GS}S_{GS}} + \beta_{sp} \frac{N_{GS}}{\tau_{sp}} \tag{11}$$

in which $N_i$ and $S_i$ respectively denote the number of carriers and photons in level i. $\tau_{sp}$ is the spontaneous recombination time, $\tau_s$ is the photon lifetime into cavity [18], $\tau_{0wl-ES_2}$ is the initial capture time to $ES_2$, and $\tau_{0ES_2-ES_1}$ and $\tau_{0ES_1-GS}$ are initial relaxation times respectively to $ES_2$ and GS. Also,

$$\tau_{ES1-GS} = \frac{\tau_{0ES_1-GS}}{1-f_{GS}}, \qquad \tau_{ES_2-ES_1} = \frac{\tau_{0ES_2-ES_1}}{1-f_{ES_1}} \tag{12}$$

We suppose these factors to be fixed when indium percentage change is infinitesimal. $\beta_{sp}$ is the spontaneous emission factor, $\eta_i$ is the injection efficiency, $\Gamma$ is the optical confinement factor, and $v_g$ is group velocity into the cavity. $f_i = N_i/\mu_i N_D$ is the occupation probability function in level $i$ with degeneracy $\mu_i$ and $N_D$ as the total number of QDs in the active region; $\epsilon_{m_i}$ is defined as the gain compression factor of level $i$

$$\epsilon_{m_i} = \frac{e^2 p_{cv}^2 \tau_s}{4\hbar n_r^2 m_0^2 \epsilon_0 E_i \Gamma_{hom}}, \tag{13}$$

with transition matrix

$$p_{cv}^2 = \frac{m_0^2 E_g(E_g+\delta)}{12 m_e\left(E_g+\frac{2\delta}{3}\right)}, \tag{14}$$

and gain factor $\epsilon_i = \frac{\epsilon_{m_i} \Gamma_{hom}}{V_a}$ in which $\Gamma_{hom}$ is the homogeneous broadening factor. Also,

$$K_i = \frac{2\pi e^2 \hbar \mu_i \xi p_{cv}^2}{c n_r \epsilon_0 m_0^2 V_{Dot} \gamma_0 E_i} \tag{15}$$

in which $\gamma_0$ describes the inhomogeneous broadening coefficient and $\xi = N_{Dot} V_{Dot}$ represents the coverage of dots with $N_{Dot}$ as the dot density and $V_{Dot}$ as the dot volume obtained as $V_{Dot} = \pi(r_{top}^2 + r_{bottom}^2)H/2$.

Here, $N_i$ and $S_i$ denote respectively the carrier and photon numbers in energy level $i$. These seven coupled differential equations have been solved simultaneously by the Runge-Kutta (IV) approach to achieve the lasing behavior by time. The values used in this paper are given in Table 2.

| Description | Symbol | Value (in SI) |
|---|---|---|
| Carrier injection rate | $\eta_i$ | $0.9$ |
| Optical confinement factor | $\Gamma$ | $0.1$ |
| Spontaneous emission coupling factor | $\beta_{sp}$ | $10^{-4}$ |
| Optical loss | $\alpha_i$ | $6 \times 10^2$ |
| Cavity's reflectivity of mirrors | $R_1, R_2$ | $R_1 = R_2 = 0.3$ |
| Decay time in wl | $\tau_{qr}$ | $5 \times 10^{-9}$ |
| Decay time in es2 es1 gs | $\tau_r$ | $5 \times 10^{-9}$ |
| Carrier capture time from wl to es2 | $\tau_c$ | $2 \times 10^{-12}$ |
| Relaxation from es2 to es1 | $\tau_{0ES_2-ES_1}$ | $2 \times 10^{-12}$ |
| Relaxation time for es1 to gs | $\tau_{0ES_1-GS}$ | $12 \times 10^{-12}$ |
| Spontaneous recombination time | $\tau_{sp}$ | $500 \times 10^{-12}$ |
| Degeneracy | $\mu_{GS}, \mu_{ES_1}, \mu_{ES_2}$ | $2, 4, 6$ |
| Number of layers | $N_l$ | $5$ |
| Cavity width | $w$ | $1 \times 10^{-5}$ |
| Active region length | $L$ | $1.5 \times 10^{-3}$ |
| Inhomogeneous broadening factor | $\gamma_0$ | $2 \times 10^{-2} e$ |
| Homogenous broadening factor | $\Gamma_{hom}$ | $10^{-2} e$ |

*Table 2: Data applied in this project. All the values are represented in SI [18, 40, 41].*

**III-B. Results and discussions on lasing**

Here the dot density is taken fixed when indium percentage varies. Fig. 4(a) shows the GS photon number in terms of time at I=7mA. As it is seen, after a turn-on delay, relaxation oscillations start which settle down to a stable radiation. The turn-on delay sounds to be small for higher amounts of indium content. This can be interpreted as result of the time required for injected carriers to distribute in the active region, relax to the GS level, and recombine with holes [12]. The difference in the turn-on delay of QDs with various amounts of indium can be explained by the fact that the energy difference between barrier and the QD energy levels is less for more indium content. Moreover, as it can be seen, amplitude of relaxation oscillations increases when indium percentage is high. In addition, photon number in stable conditions is a bit less in less indium percentage.

Also, in Figs 4(b) and 4(c) photon number is shown respectively for lasing from $ES_1$ and $ES_2$. These levels have energy of more than GS, and thus more current is needed for lasing from $ES_1$ and $ES_2$, since firstly lower levels use the current for lasing. The threshold current was found to be ~0.75 A for $ES_1$ and ~5.25 A for $ES_2$. As it is seen, general behavior of stable photon numbers, turn-on delay, and amplitude of relaxation oscillations contradicts with GS ones; by indium increase, turn-on delay, amplitude of relaxation oscillations, and number of photons in the steady state increase. This is inversely true for ES levels.

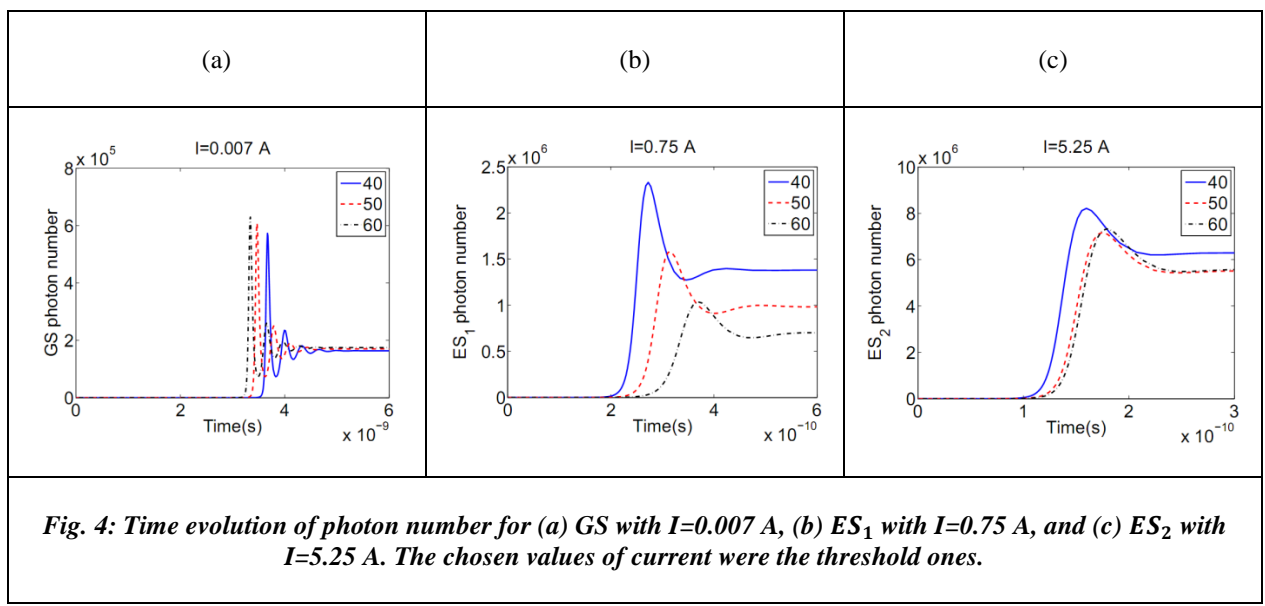

*Fig. 4: Time evolution of photon number for (a) GS with I=0.007 A, (b) $ES_1$ with I=0.75 A, and (c) $ES_2$ with I=5.25 A. The chosen values of current were the threshold ones.*

Moreover, Figs 5(a)-5(c) compare the small signal modulation response as function of modulation frequency [10, 42] under different ratios of indium content calculated for GS, $ES_1$ and $ES_2$. Here a small pulse signal stimulates the laser; It is seen that larger ratio of indium in the QD leads to larger 3dB modulation bandwidth in GS, but this is inverse for ES levels.

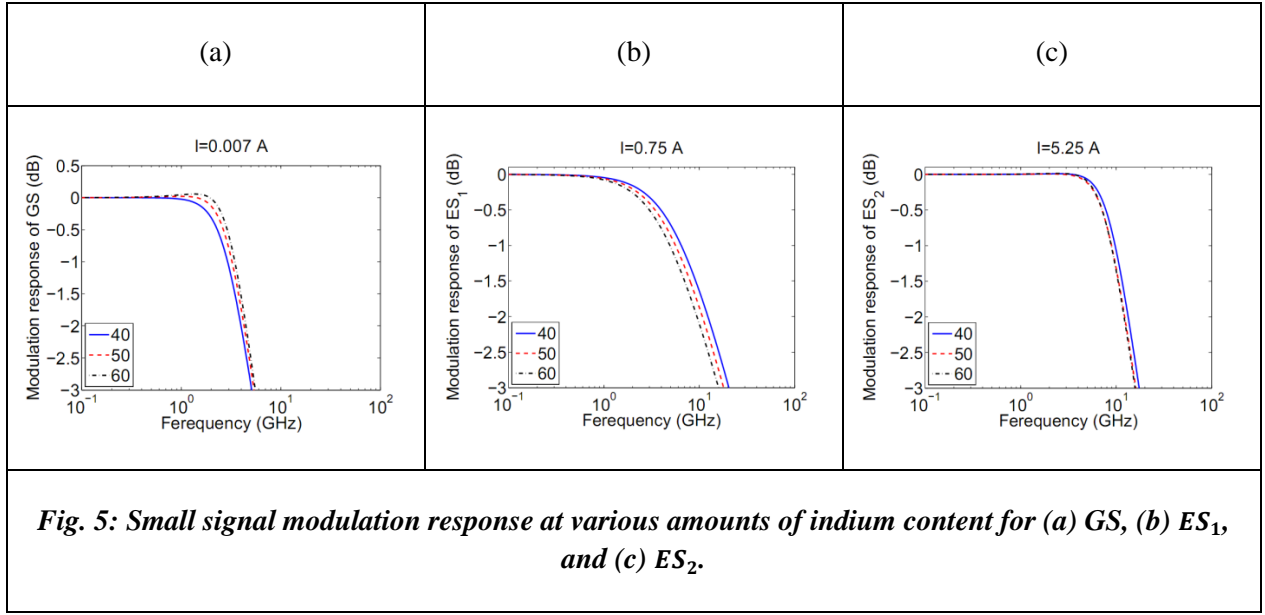

Fig. 5: Small signal modulation response at various amounts of indium content for (a) GS, (b) $ES_1$, and (c) $ES_2$.

Output power from level $i$ is calculated as follows:

$$P_{out_i} = \frac{cE_i S_i \log\left(\frac{1}{R}\right)}{2 n_r L} \quad (16)$$

where R is reflectivity index, $c$ is the light speed, $L$ is the cavity length, and $n_r$ is cavity refractive index. This parameter is plotted in stable lasing versus current in Fig. 6 for GS, $ES_1$ and $ES_2$ for three different indium amounts. As it is seen, threshold current is independent of indium percentage, and this is true for all levels. Moreover, power is greater for QDs with less indium amount.

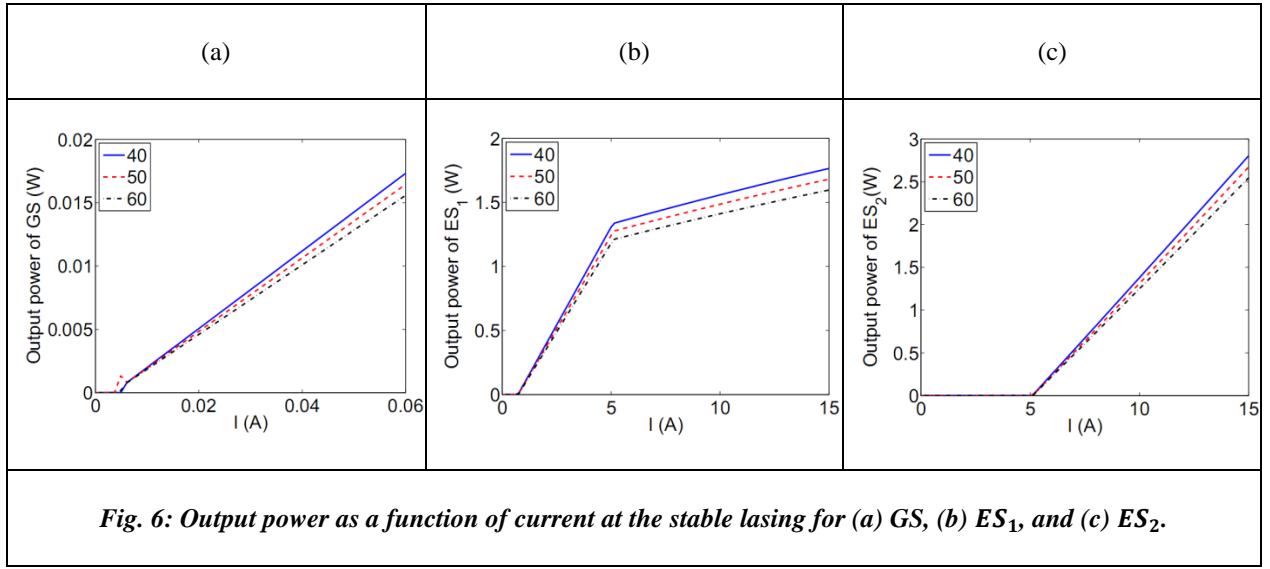

Fig. 6: Output power as a function of current at the stable lasing for (a) GS, (b) $ES_1$, and (c) $ES_2$.

Optical gain of level $i$ is as follows:

$$Gain_i = \frac{\Gamma K_i\left(\frac{2N_i}{\mu_i N_D}-1\right)}{1+\epsilon_i S_i} \qquad (17)$$

which is represented in Figures 7(a)-6(c) as a function of injected current for all energy states. As it is observed, change of indium content has no remarkable effect on the laser gain.

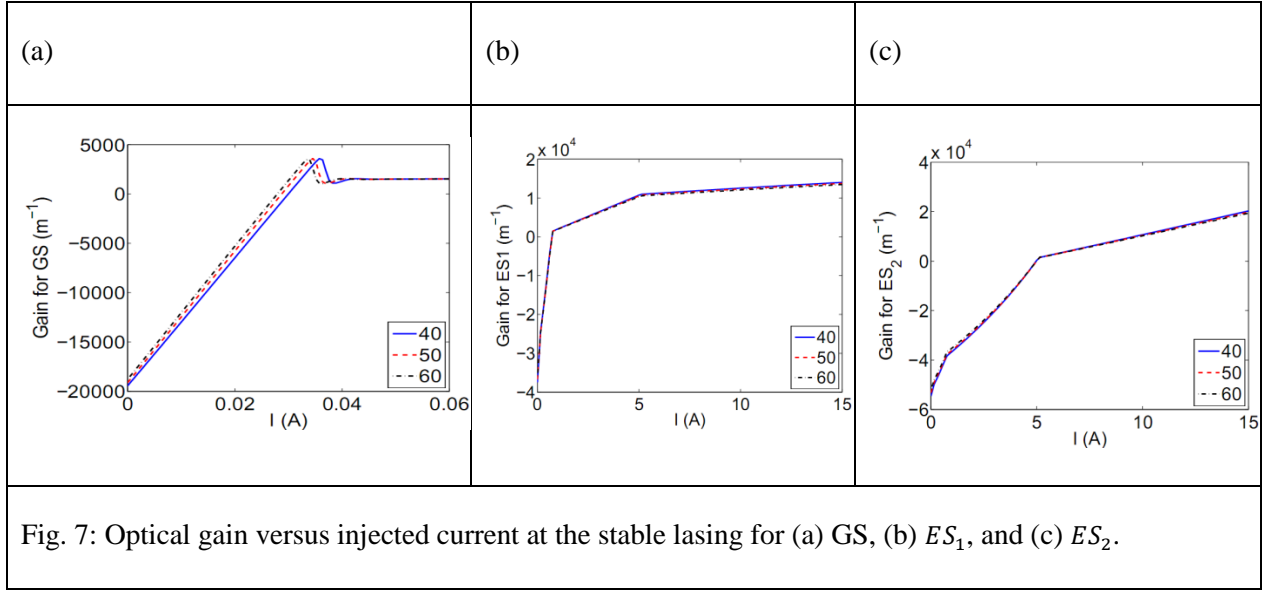

Fig. 7: Optical gain versus injected current at the stable lasing for (a) GS, (b) $ES_1$, and (c) $ES_2$.

Fig. 8 exhibits also the dependence of output power to cavity length. As it is seen, injected current is 0.1, 2 and 15 A respectively for GS, $ES_1$, and $ES_2$. As it is seen, a peak is observed at all figures, showing a maximum power for a special cavity length. This optimized cavity length and the related output power were found to be strongly sensitive to the injected current. However, as it is seen, change of indium percentage is not a determinant, as it only changes the value of maximum power.

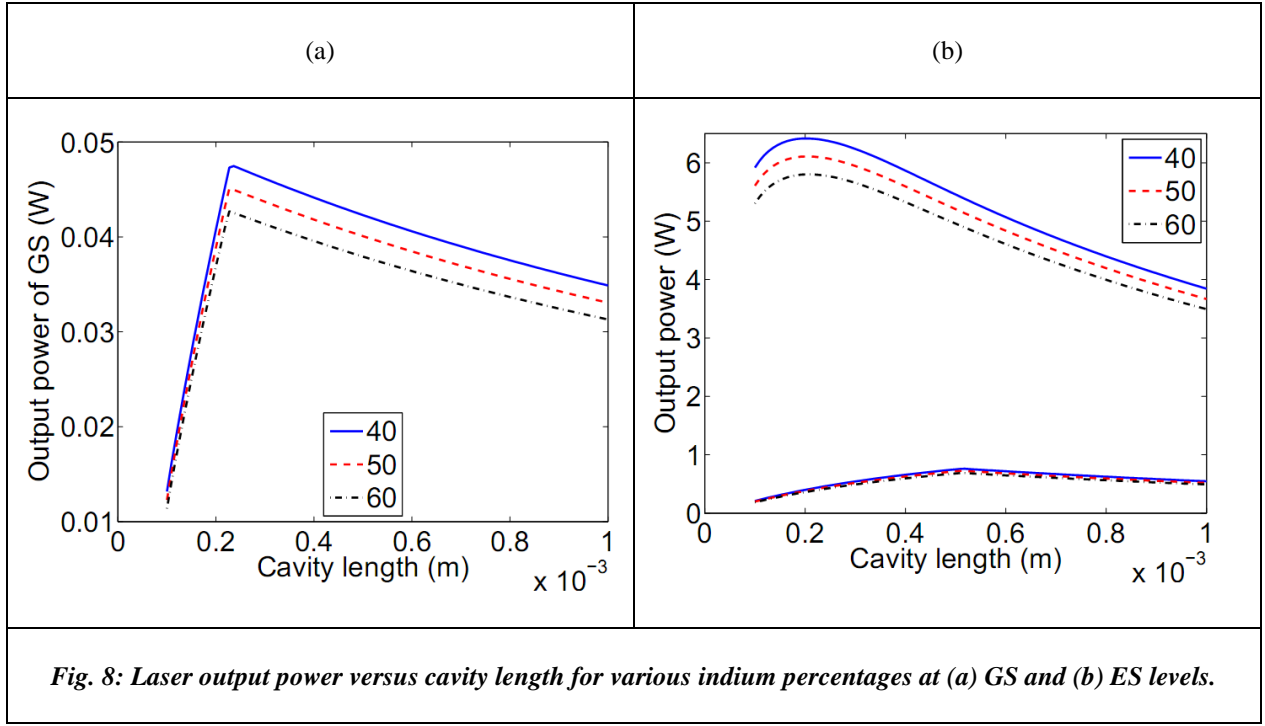

*Fig. 8: Laser output power versus cavity length for various indium percentages at (a) GS and (b) ES levels.*

## IV. CONCLUSION

We surveyed the influence of indium percentage on laser properties in $In_xGa_{1-x}As/GaAs(001)$ lasers. Energy levels of truncated-cone-shape QDs were calculated by k.p model. Then, by introducing a three-level model, laser outputs were examined. Our findings indicate that inclusion of more indium leads to reduction of energy gap and e-h recombination energy. In addition, laser output for both GS and ES levels was dependent on indium percentage. It was shown that indium percentage increase at fixed injected current results in the increased GS photon number, ES turn-on delay, and GS 3dB modulation bandwidth and decreased GS turn-on delay, ES photon number, amplitude of relaxation oscillations, output power, and ES 3dB modulation bandwidth; but has no effect on threshold current and laser gain. At last, we found an optimized cavity length which was likely to be independent of indium percentage.

**Acknowledgement**

The authors give the sincere appreciation to Dr. S. Birner for providing the advanced 3D Nextnano++ simulation program [43] and his instructive guides. We also thank Prof. S. Farjami Shayesteh for comments on the manuscript.

**References:**

1. Markéta ZÍKOVÁ, A.H., *Simulation of Quantum States in InAs/GaAs Quantum Dots.* NANOCON 2012. **23**(25): p. 10.
2. Ma, Y.J., et al., *Factors influencing epitaxial growth of three-dimensional Ge quantum dot crystals on pit-patterned Si substrate.* Nanotechnology, 2013. **24**(1): p. 015304.
3. Danesh Kaftroudi, Z. and E. Rajaei, *Simulation And Optimization Of Optical Performance Of Inp-Based Longwavelength Vertical Cavity Surface Emitting Laser With Selectively Tunnel Junction Aperture.* Journal Of Theoretical And Applied Physics (Iranian Physical Journal), 2010. **4**(2): p. 12-20.
4. Nedzinskas, R., et al., *Polarized photoreflectance and photoluminescence spectroscopy of InGaAs/GaAs quantum rods grown with As(2) and As(4) sources.* Nanoscale Research Letters, 2012. **7**(1): p. 609-609.
5. Costantini, G., et al., *Interplay between Thermodynamics and Kinetics in the Capping of $\mathrm{InAs}/\mathrm{GaAs}(001)$ Quantum Dots.* Physical Review Letters, 2006. **96**(22): p. 226106.
6. Bimberg, D., et al., *Quantum dot lasers: breakthrough in optoelectronics.* Thin Solid Films, 2000. **367**(1–2): p. 235-249.
7. Gioannini, M., *Analysis of the Optical Gain Characteristics of Semiconductor Quantum-Dash Materials Including the Band Structure Modifications Due to the Wetting Layer.* IEEE Journal of Quantum Electronics, 2006. **42**(3): p. 331-340.
8. Danesh Kaftroudi, Z. and E. Rajaei, *Thermal simulation of InP-based 1.3 micrometer vertical cavity surface emitting laser with AsSb-based DBRs*. Vol. 284. 2011, Amsterdam, PAYS-BAS: Elsevier. 11.
9. Asryan, L.V. and S. Luryi, *Tunneling-injection quantum-dot laser: ultrahigh temperature stability.* Quantum Electronics, IEEE Journal of, 2001. **37**(7): p. 905-910.
10. Xu, P.-F., et al., *Temperature-dependent modulation characteristics for 1.3µm InAs/GaAs quantum dot lasers.* Journal of Applied Physics, 2010. **107**(1): p. 013102.
11. Martin, H., et al., *In(Ga)As quantum dot formation on group-III assisted catalyst-free InGaAs nanowires.* Nanotechnology, 2011. **22**(19): p. 195601.
12. Yekta Kiya, Y., E. Rajaei, and A. Fali, *Study of response function of excited and ground state lasing in InGaAs/GaAs quantum dot laser.* J. Theor. Phys. , 2012. **1**: p. 246-256.
13. Shafieenezhad, A., E. Rajaei, and S. Yazdani, *The Effect of Inhomogeneous Broadening on Characteristics of Three-State Lasing InGaAs/GaAs Quantum Dot Lasers.* International Journal of Scientific Engineering and Technology, 2014. **3**(3): p. 297- 301.
14. Yazdani, S., E. Rajaei, and A. Shafieenezhad, *Optimizing InAs/InP (113) B quantum dot lasers with considering mutual effects of coverage factor and cavity length on two-state lasing.* International Journal of Engineering Research, 2014. **3**(3): p. 172-176.
15. Dieter, B., *Quantum dots for lasers, amplifiers and computing.* Journal of Physics D: Applied Physics, 2005. **38**(13): p. 2055.
16. Razm-Pa, M. and F. Emami, *Effect of parameter variations on the static and dynamic behaviour of a self-assembled quantum-dot laser using circuit-level modelling.* Quantum Electronics, 2015. **45**(1): p. 15.
17. Hazdra, P., et al., *Optical characterisation of MOVPE grown vertically correlated InAs/GaAs quantum dots.* Microelectronics Journal, 2008. **39**(8): p. 1070-1074.
18. Fali, A., E. Rajaei, and Z. Kaftroudi, *Effects of the carrier relaxation lifetime and inhomogeneous broadening on the modulation response of InGaAs/GaAs self-assembled quantum-dot lasers.* Journal of the Korean Physical Society, 2014. **64**(1): p. 16-22.

19. Yu, L.K., et al., *The effect of In content on high-density InxGa1–xAs quantum dots.* Journal of Crystal Growth, 2005. **282**(1-2): p. 173-178.
20. Shrestha, S.K., et al., *Accurate stoichiometric analysis of polycrystalline indium nitride films with elastic recoil detection.* Current Applied Physics, 2004. **4**(2–4): p. 237-240.
21. Shi, Z., et al., *Influence of V/III ratio on QD size distribution.* Frontiers of Optoelectronics in China, 2011. **4**(4): p. 364-368.
22. Maia, A.D.B., et al., *The influence of different indium-composition profiles on the electronic structure of lens-shaped In x Ga 1– x As quantum dots.* Journal of Physics D: Applied Physics, 2012. **45**(22): p. 225104.
23. Shahraki, M. and E. Esmaili, *Computer simulation of quantum dot formation during heteroepitaxial growth of thin films.* Journal of Theoretical and Applied Physics, 2012. **6**(1): p. 1-5.
24. Aryanto, D., Z. Othaman, and A.K. Ismail, *The impact of AsH3 overflow time and indium composition on the formation of self-assembled In x Ga1 – x As quantum dots studied by atomic force microscopy.* Journal of Theoretical and Applied Physics, 2013. **7**(27): p. 1-6.
25. Kamath, K., et al., *Small-signal modulation and differential gain of single-mode self-organized In0.4Ga0.6As/GaAs quantum dot lasers.* Applied Physics Letters, 1997. **70**(22): p. 2952-2953.
26. Stracke, G., et al., *Indirect and direct optical transitions in In0.5Ga0.5As/GaP quantum dots.* Applied Physics Letters, 2014. **104**(12): p. 123107.
27. Jun, Y., P. Bhattacharya, and Z. Mi, *High-Performance In$_{0.5}$Ga$_{0.5}$As/GaAs Quantum-Dot Lasers on Silicon With Multiple-Layer Quantum-Dot Dislocation Filters.* Electron Devices, IEEE Transactions on, 2007. **54**(11): p. 2849-2855.
28. Zieliński, M., M. Korkusiński, and P. Hawrylak, *Atomistic tight-binding theory of multiexciton complexes in a self-assembled InAs quantum dot.* Physical Review B, 2010. **81**(8): p. 085301.
29. Korkusinski, M., M. Zielinski, and P. Hawrylak, *Multiexciton complexes in InAs self-assembled quantum dots.* Journal of Applied Physics, 2009. **105**(12): p. 122406.
30. Qiu, D. and M.X. Zhang, *The preferred facet orientation of GaAs pyramids for high-quality InAs and InxGa1–xAs quantum dot growth.* Scripta Materialia, 2011. **64**(7): p. 681-684.
31. Trellakis, A., et al., *The 3D nanometer device project nextnano: Concepts, methods, results.* Journal of Computational Electronics, 2006. **5**(4): p. 285-289.
32. Jang, Y.D., et al., *Comparison of quantum nature in InAs/GaAs quantum dots.* Journal of the Korean Physical Society, 2003. **42**(Suppl): p. 111-113.
33. Singh, J., *Physics of Semiconductors and Their Heterostructures*. 1993: McGraw-Hill.
34. Yu, C., *Fundamentals of Semiconductors.* Springer, 2010.
35. Adachi, S., J.Appl. Phys., 1983. **vol. 54**(No.4): p. 1844-1848.
36. T.P.Pearsall, *GaInAsP Alloy Semiconductors.* John Wiley and Sons, 1982.
37. N.M., G.Y.A.a., *Schmidt Handbook Series on Semiconductor Parameters.* World Scientific, London, 1999. **vol. 2**: p. 62-88.
38. Bratkovski, A. and T.I. Kamins, *Nanowire-Based Light-Emitting Diodes and Light-Detection Devices With Nanocrystalline Outer Surface*. 2010, Google Patents.
39. Goetz, K.H., et al., *Optical and crystallographic properties and impurity incorporation of GaxIn1−xAs (0.44<x<0.49) grown by liquid phase epitaxy, vapor phase epitaxy, and metal organic chemical vapor deposition.* Journal of Applied Physics, 1983. **54**(8): p. 4543-4552.
40. Lv, S.-f., et al., *Modeling and simulation of InAs/GaAs quantum dot lasers.* Optoelectronics Letters, 2011. **7**(2): p. 122-125.
41. Gioannini, M., *Ground-state power quenching in two-state lasing quantum dot lasers.* Journal of Applied Physics, 2012. **111**(4): p. 043108.

42. Horri, A. and R. Faez, *Small signal circuit modeling for semiconductor self-assembled quantum dot laser.* Optical Engineering, 2011. **50**(3): p. 034202-034202-5.
43. Birner, S., et al., *nextnano: General Purpose 3-D Simulations.* Electron Devices, IEEE Transactions on, 2007. **54**(9): p. 2137-2142.